\input graphicx

\def\a{\alpha}\def\b{\beta}\def\d{\delta}
\def\g{\gamma}
\def\l{\lambda}\def\m{\mu}\def\n{\nu}\def\o{\omega}\def\q{\psi}\def\r{\rho}\def\s{\sigma}
\def\y{\eta}

\def\D{\Delta}

\def\coo{coordinates }\def\des{de Sitter }\def\ads{(anti)-de Sitter }
\def\cor{commutation relations }\def\ur{uncertainty relation }
\def\mn{{\mu\nu}}\def\lra{\leftrightarrow}\def\bdot{\!\cdot\!}
\def\ie{i.e.\ }\def\({\left(}\def\){\right)}\def\cc{coupling constant }
\def\schr{Schr\"odinger }\def\de{\partial}\def\bc{boundary conditions }
\def\ha{{1\over 2}}\def\rep{representation }\def\inf{\infty}

\def\section#1{\bigskip\noindent{\bf#1}\smallskip}

\def\subsection#1{\smallskip\noindent{\it#1}\smallskip}

\font\small = cmr8

\def\PL#1{Phys.\ Lett.\ {\bf#1}}

\def\PR#1{Phys.\ Rev.\ {\bf#1}}\def\CQG#1{Class.\ Quantum Grav.\ {\bf#1}}
\def\NP#1{Nucl.\ Phys.\ {\bf#1}}\def\GRG#1{Gen.\ Relativ.\ Grav.\ {\bf#1}}
\def\JMP#1{J.\ Math.\ Phys.\ {\bf#1}}
\def\PRS#1{Proc.\ R. Soc.\ Lond.\ {\bf#1}}
\def\JoP#1{J.\ Phys.\ {\bf#1}} \def\IJMP#1{Int.\ J. Mod.\ Phys.\ {\bf #1}}
\def\MPL#1{Mod.\ Phys.\ Lett.\ {\bf #1}}

\def\RMP#1{Rev.\ Mod.\ Phys.\ {\bf#1}}

\def\arx#1{{\tt arXiv:#1}}

\def\ref#1{\medskip\everypar={\hangindent 2\parindent}#1}
\def\beginref{\begingroup
\bigskip
\centerline{\bf References}
\nobreak\noindent}
\def\endref{\par\endgroup}

\def\hx{{\hat x}}\def\hp{{\hat p}}\def\hb{\hbar}

{\nopagenumbers
\line{}
\vskip40pt
\centerline{\bf QUANTUM MECHANICS OF THE NONRELATIVISTIC YANG MODEL}
\vskip50pt
\centerline{{\bf S. Meljanac}\footnote{$^\dagger$}{e-mail: meljanac@irb.hr}}
\vskip5pt
\centerline {Rudjer Bo\v skovi\'c Institute, Theoretical Physics Division}
\centerline{Bljeni\v cka c.~54, 10002 Zagreb, Croatia}
\vskip10pt
\centerline{{\bf S. Mignemi}\footnote{$^\ast$}{e-mail: smignemi@unica.it}}
\vskip5pt
\centerline {Dipartimento di Matematica, Universit\`a di Cagliari}
\centerline{via Ospedale 72, 09124 Cagliari, Italy}
\smallskip
\centerline{and INFN, Sezione di Cagliari}
\centerline{Cittadella Universitaria, 09042 Monserrato, Italy}

\vskip80pt
\centerline{\bf Abstract}
\medskip
{\noindent We discuss, at leading order in $\hbar$, the quantum mechanics of a specific realization in phase space of the Yang model
describing noncommutative geometry in a curved background.
In particular, we show how the deformation of the Heisenberg \ur crucially depends on the signs of the coupling
constants of the model. We also discuss the dynamics of the free particle and of the harmonic oscillator. Also in this case
the results depend on the signs of the coupling constants.
}
\vskip60pt
\vfil\eject}
\section{1. Introduction}
The Yang model [1] was introduced in 1947 as a generalization of the Snyder model [2] of noncommutative geometry to a curved
spacetime background.
Its aim was to describe the symmetries of a quantum phase space through an algebra
which includes position and momentum \coo $\hx$ and $\hp$, together with Lorentz generators and a further scalar generator,
necessary to close the algebra. The algebra turns out to be isomorphic to $o(1,5)$, and
includes as subalgebras the Snyder and the \des algebra.
The Yang model can therefore be interpreted as describing a noncommutative geometry in a spacetime of constant curvature.
An interesting property of the model is its duality for the exchange of position and
momentum coordinates, together with the parameters $\a$ and $\b$ (see below).
This recalls the old Born duality proposal [3].

The Yang model can also be interpreted as a deformation of the standard symplectic structure of flat-space quantum mechanics by
two parameters $\a$ and $\b$, that curve spacetime and momentum space [4], inducing a noncommutative structure on a \des space.
The two parameters can have both positive and negative values, giving rise to very different physical properties: positive $\a$
corresponds to the symmetries of de Sitter spacetime, while negative $\a$ to those of \ads spacetime.
On the other hand, $\b>0$ generates the Snyder algebra, that implies the existence of a maximal mass, while $\b<0$ yields the
so-called anti-Snyder algebra, which does not entail such bound [5]. For $\a=\b=0$ one recovers the standard theory.

To give an interpretation to the physics associated to the Yang algebra, it is natural to identify $\a$ and $\b$
with the cosmological constant and the inverse of the Planck mass squared, respectively. The parameters are therefore
exceedingly small for experimental scales related to particle physics and gravity.
Although from a mathematical point of view all values of positions and momenta are allowed (except that for positive
values of the coupling constants $x^2<1/\a$ and $p^2<1/\b$),
the theory is supposed to be reliable only for values of $x^2\ll1/\a$ and $p^2\ll1/\b$.

Although the Yang model has attracted growing interest recently, only its formal properties have been investigated [6],
with the exception of ref.~[7], where some implications on statistical mechanics and on the uncertainty principle were considered.
Moreover, connections with different theories, like conformal field theory or fuzzy gravity, have been proposed by
some authors [8].

In this paper we shall try to extract some physical consequences of the formalism in simple situations.
For this purpose, it is useful to consider the nonrelativistic version of the model, defined on a three-dimensional Euclidean
space. In fact, the discussion of the nonrelativistic model is easier, in particular if one
is interested in the study of quantum mechanics on a space with Yang geometry, since a relativistic version of
quantum mechanics presents several problems, which are usually argued to be resolved only by quantum field theory.

In this framework, we can discuss the modification of the \ur induced by the deformed symplectic structure.
Deformed \ur have earned much interest in the last years, since they may emerge in a variety of situations, including string theory,
quantum gravity, noncommutative geometry or gedanken experiments and are strictly related to the existence of a minimal length [9].
It must be remarked, however, that not all deformed uncertainty relations predict the existence of a minimal length [10]. As we shall
see, this property is related to the sign of the correction to the Heisenberg relations.

However, also the curvature of spacetime leads to a deformation of the \ur [11].
Both deformation can be unified in the so called Extended Generalized Uncertainty Principle (EGUP) [12,13].
Of course, as we shall show, it is this last case that occurs in the Yang model.

In the context of deformed uncertainty relations, also the investigation of simple quantum mechanical models,
like the free particle or the harmonic oscillator, is interesting and has been performed by several authors [14,15]. In particular,
it has been shown that in some cases the free particle has discrete spectrum, while the harmonic oscillator always displays corrections
with respect to the standard case. More recently, more realistic models have been studied in the context of molecular dynamics,
as for example diatomic molecules [16].

The results we shall obtain for the Yang model can be compared with those of a model based on similar assumption, that was introduced
much later with the name of Triply Special Relativity (TSR) or Snyder-de Sitter (SdS) [17].
In such model the same $o(1,5)$ algebra is realized in a nonlinear way, so that the additional scalar generator is no longer
necessary. This fact simplifies the physical interpretation of the theory and its realization on phase space.
The modified \ur and the deformed spectrum of the harmonic oscillator for the nonrelativistic version of this model were discussed
in [18].

In the present case the investigation of the deformations of the \ur and of the dynamics can be done only if one considers a specific
realization of the model on phase space, since, as noted above, it is otherwise
difficult to give a physical interpretation to the scalar generator of the algebra.
Moreover, to avoid computational complications, we shall limit our study to the one-dimensional case and to the leading order in $\hbar$,
$\a$ and $\b$.
We shall consider both positive and negative values of the coupling parameters, since they lead to very different physical predictions.

By comparing our results with the known case of TSR, we shall see that they share some qualitative similarities, but
also differ in important respects.

We shall give a heuristic derivation of our results. A rigorous mathematical treatment, including for example the discussion of the
functional analysis of the position and momentum operators, would require much more details, and we plan to expose it in a forthcoming
paper.
\bigbreak

\section{2. Realization of the Yang model on canonical phase space}
In detail, the Yang algebra is defined by the \cor
$$\eqalignno{&[\hx_\m,\hx_\n]=i\hb\b M_\mn,\qquad[\hp_\m,\hp_\n]=i\hb\a M_\mn,\qquad[\hx_\m,\hp_\n]=i\hb K\y_\mn,\cr
&[K,\hx_\m]=i\hb\b\hp_\m,\qquad[K,\hp_\m]=-i\hb\a\hx_\m,\qquad[M_\mn,K]=0,\cr
&[M_\mn,\hx_\l]=i\hb(\y_{\m\l}\hx_\n-\y_{\n\l}\hx_\m),\qquad[M_\mn,\hp_\l]=i\hb(\y_{\m\l}\hp_\n-\y_{\n\l}\hp_\m),\cr
&[M_\mn,M_{\r\s}]=i\hb\big(\y_{\m\r}M_{\n\s}-\y_{\m\s}M_{\n\r}-\y_{\n\r}M_{\m\s}+\y_{\n\s}M_{\m\r}\big),&(1)}$$
where $\a$ and $\b$ are real parameters and $\y_\mn$ the flat metric.

Clearly, its properties depend on the signs of the parameters $\a$ and $\b$. In fact, the Yang algebra  is isomorphic to
$o(1,5)$, $o(2,4)$ or $o(3,3)$, depending on these signs, and the physics varies accordingly, as in the similar case of TSR [18].
However, contrary to TSR, in our case it is not necessary that the two parameters have the same sign.

We interpret the operators $\hx_\m$ and $\hp_\m$ as \coo of the quantum phase space, $M_\mn$ as generators of
the Lorentz transformations and $K$ as a further generator, necessary to close the algebra.
The algebra (1) is invariant under a generalized Born duality, $\a\lra\b$, $\hx_\m\to\hp_\m$,  $\hp_\m\to-\hx_\m$,
$M_\mn\lra M_\mn$, $K\lra K$.
In the limit $\b\to0$ it describes the phase-space symmetries of the (anti)-de Sitter spacetime, in the limit $\a\to0$ those of
the (anti)-Snyder one.

In order to give a physical interpretation of the Yang model,
it is useful to find a realization of its algebra on a canonical quantum phase space, with
$$[x_\m,x_\n]=[p_\m,p_\n]=0,\qquad[x_\m,p_\n]=i\hb\y_\mn,\eqno(2)$$

In [19] it has been shown that the simplest realization  is given to leading order in $\hbar$ (\ie in the classical limit),
by
$$\hx_\mu= x_\mu\sqrt{\sqrt{1+\a\b(x\bdot p)^2}-\b p^2}\,\qquad \hp_\mu=p_\mu\sqrt{\sqrt{1+\a\b(x\bdot p)^2}-\a x^2}\ ,\eqno(3)$$
with
$$M_\mn=x_\m p_\n-x_\n p_\m={\hx_\m\hp_\n-\hx_\n\hp_\m\over K},\eqno(4)$$
and
$$K=\sqrt{1-\a\hx^2-\b\hp^2-{\a\b\over2}M_{\r\s}^2}=\sqrt{1-\a\hx^2-\b\hp^2-\a\b\,{\hx^2\hp^2-(\hx\bdot\hp)^2\over K^2}}.\eqno(5)$$
From (3) it follows that if $\a\b<0$, the scalar product $x\bdot p$ must have an upper bound $1/\sqrt{|\a\b|}$, while this is not
true if $\a\b>0$.

In terms of the original variables $\hx_\m$ and $\hp_\m$, equation (5) yields
$$K=\sqrt{{1-\a\hx^2-\b\hp^2\over2}\(1+\sqrt{1-4\a\b{\hx^2\hp^2-(\hx\bdot\hp)^2\over(1-\a\hx^2-\b\hp^2)^2}}\ \)}.\eqno(6)$$

Starting from these results, one can investigate the 3D Euclidean version of the model. As  discussed in the Introduction, this is interesting because
in this way one can define a nonrelativistic quantum mechanics based on the Yang \cor and obtain for example deformed uncertainty relations.
We call the ensuing model nonrelativistic Yang model. Notice however that, as noticed in analogous cases [20], this is not to be identified with the
$c\to\inf$ limit of the Yang model.
The nonrelativistic version of TSR has been investigated for example in [18], while for the Snyder model, it has been studied in [5].

In the nonrelativistic model, the \cor are the same as (1), but now the indices run from 1 to 3, $\y_\mn\to\d_\mn$ and the algebra reduces to $o(5)$,
$o(1,4)$ or $o(2,3)$, depending on the signs of $\a$ and $\b$.

\bigbreak
\section{3. Generalized uncertainty principle}
A discussion of the properties of the Yang model in the realization of the previous section is however extremely difficult, even in the 3D Euclidean  version,
but the problem is highly simplified in the one-dimensional limit. Although in one dimension one cannot speak of noncommutativity and curvature
of spacetime, this limit has nevertheless proven to maintain the most interesting features of its higher-dimensional counterparts in similar models
[18], since its behaviour essentially coincides with that of the higher-dimensional radial equation, corresponding to the zero angular momentum sector.
In the Snyder case this limit has also been widely investigated in the context of minimal length models [14,15]. Recently, the appearance of deformed \ur in
the 1D Yang model has independently been observed in [7].

In our approximation, the one-dimensional commutation relation for $\hx$ and $\hp$ , with $K$ given by (5), simplifies to
$$[\hx,\hp]=i\hbar\sqrt{1-\a\hx^2-\b\hp^2}.\eqno(7)$$
It follows that for $\a>0$, the acceptable values of $\hx^2$ are bounded by $1/\a$, and for $\b>0$ those of $\hp^2$ by $1/\b$, while
no such bound is instead present for negative values of the coupling constants.
Moreover, in general one must require $\a\hx^2+\b\hp^2\le1$.

Assuming for simplicity $\langle\hx\rangle=\langle\hp\rangle=0$, by the Robinson-Schr\"odinger inequality, the commutation relation (7) implies a
generalized uncertainty principle, analogous to EGUP,\footnote{$^1$}{Note that in most cases studied in the literature the corrections are defined with
opposite sign.}
$$\D\hx\D\hp\ge\ha\left|\langle[\hx,\hp]\rangle\right|\approx{\hbar\over2}\(1-{\a\over2}(\D\hx)^2-{\b\over2}(\D\hp^2)\),\eqno(8)$$
where we have expanded at first order in the parameters $\a$ and $\b$. Mathematically, the expansion is valid for $\hx^2<1/\a$ if $\a>0$ and
$\hp^2<1/\b$ if $\b>0$, while it is always valid for negative values of the parameters. Moreover, considering only first-order terms is justified if
$\hx^2\ll1/\a$ and $\hp^2\ll1/\b$, which is always true for physically reasonable values of the coupling constants.

Taking the square of (8) gives
$$(\D\hx)^2(\D\hp)^2\ge{\hbar^2\over4}\Big(1-\a(\D\hx)^2-\b(\D\hp)^2\Big).\eqno(9)$$
It follows that the uncertainty must satisfy
$$(\D\hx)\ge\hbar\sqrt{{1-\b(\D\hp)^2\over\hbar^2\a+4(\D\hp)^2}},\qquad(\D\hp)\ge\hbar\sqrt{{1-\a(\D\hx)^2\over\hbar^2\b+4(\D\hx)^2}}.\eqno(10)$$
The behaviour of $\D x$ in function of $\D p$ in the various cases is reported in fig.~1.

For $\a,\b>0$, the uncertainties satisfy the bounds $0\le\D\hp\le 1/\sqrt\b$, and $0\le\D\hx\le 1/\sqrt\a$, as is also obvious because of the bounds imposed
on $\hx$ and $\hp$.

For $\a,\b<0$, $\D\hp\ge\hbar\sqrt{|\a|}/2$, and $\D\hx\ge\hbar\sqrt{|\b|}/2$, hence both position and momentum have a minimal uncertainty independent of that
of the conjugated variable.


A more complicated behaviour occurs if $\a$ and $\b$ have different signs. If $\a<0$ and $\b>0$, $\D\hp$ must satisfy the bounds $\hbar\sqrt{|\a|}/2\ge\D\hp\ge1/\sqrt\b$,
while $\D\hx$ can take any positive value.

Finally, if $\a>0$ and $\b<0$, $\D p$ can take any positive value, while $\hbar\sqrt{|\b|}/2\ge\D x\ge1/\sqrt\a$.
\medskip
It is easy to check that in all cases the minimal values so obtained satisfy the bound $1-\a\hx^2-\b\hp^2\ge0$.
It is also clear that if either $\a$ or $\b$ is positive, it is possible to have states with vanishing $\D\hx\D\hp$, while this cannot occur if both
$\a$ and $\b$ are negative. It seems therefore that only the case of negative $\a$ and $\b$ can satisfy the standard uncertainty bound. However, also for
positive values of the \cc states with vanishing uncertainty are achieved only for values of $\D\hx$ and $\D\hp$ close to the limit values of the phase space
variables, which are assumed to be out of the range of applicability of the theory.

Of course, our derivation has been heuristic. A rigorous discussion would require a definition of squeezed states on a suitable Hilbert space,
in order to define maximally localized states [14,21,12]. This will be discussed in detail in a forthcoming paper.

\bigskip

\centerline{\includegraphics[width=40ex]{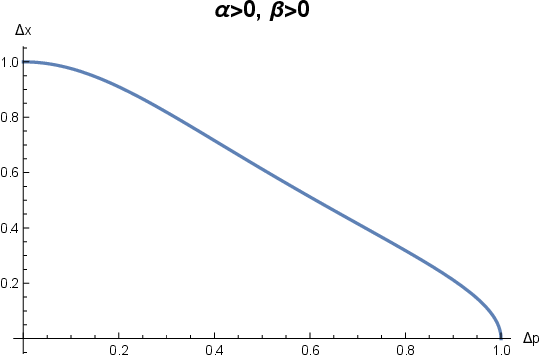}\qquad\qquad\includegraphics[width=40ex]{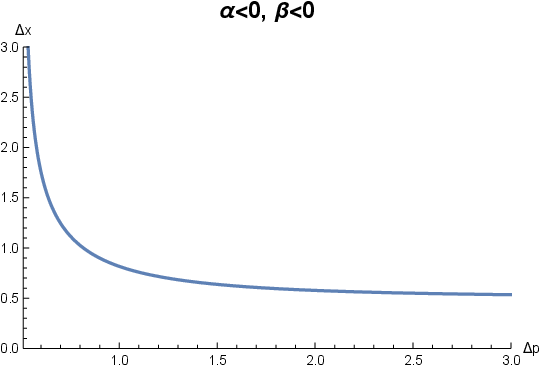}}
\bigskip
\bigskip

\centerline{\includegraphics[width=40ex]{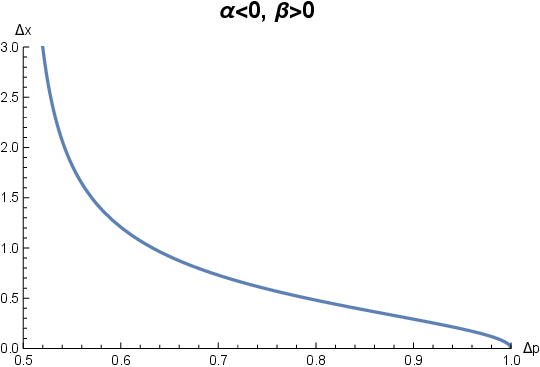}\qquad\qquad\includegraphics[width=40ex]{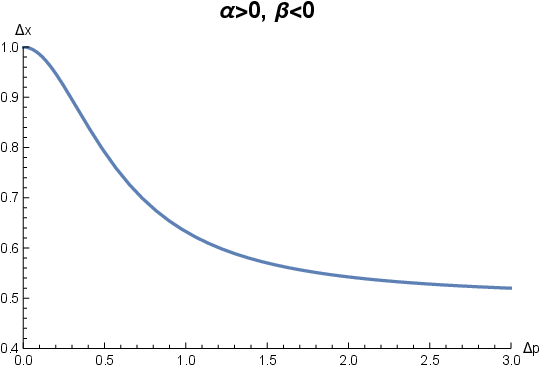}}
\bigskip

{\noindent \small Fig.~1: The minimal value of $\scriptstyle{\D x}$ in function of $\scriptstyle{\D p}$ for different signs of the coupling constants.
In the graphs, $\scriptstyle{\D x}$ and $\scriptstyle{\D p}$ are multiplied by $\scriptstyle{\sqrt{|\a|}}$ and $\scriptstyle{\sqrt{|\b|}}$, respectively.}
\bigskip

We can compare the behaviour of the \ur with that of the TSR model. We recall that in TSR $\a$ and $\b$ must have the same sign in order to preserve the
reality of the algebra. In that case the uncertainty relation is given by [18]
$$\D\hx\D\hp\ge{\hbar\over2}\,{1+\a\D\hx^2+\b\D\hp^2\over1+\hbar\sqrt{\a\b}}.\eqno(11)$$

Notably, in  TSR one has minimal values both for $\D x$ and $\D p$ when $\a,\b>0$, while if $\a,\b<0$ no minimal values arise.
The difference with the Yang model is due to the opposite sign of the quadratic terms in (7) and (11).
In TSR the minimal values of the uncertainties are given by
$$(\D\hx^2)_{min}= {\hbar^2\b\over1+2\hbar\sqrt{\a\b}},\qquad(\D\hp)^2_{min}={\hbar^2\a\over1+2\hbar\sqrt{\a\b}}.\eqno(12)$$
These are similar to the ones obtained above for the Yang model.

\section{4. One-dimensional dynamics}

Clearly, the quantum dynamics of the Yang model differs from the standard one, because of the deformation of the commutation relations.
In one dimension, the dynamics of the free particle and of the harmonic oscillator  has been studied for the Snyder model (also in the context of
the generalized uncertainty principle) [14,15,5] and TSR [18], displaying a discrete spectrum for the free particle with positive coupling constants,
and a deformation of the energy spectrum of the harmonic oscillator.

Let us then consider the dynamics of these models in the Yang case. For simplicity, we shall limit our study to the case of positive $\a$ and $\b$.
We recall that in this case $|x|<1/\a$, $|p|<1/\b$, but no minimal uncertainties occur.
We start by considering the solution of the one-dimensional \schr equation for a free particle, using the realization (3).

Since the equation obtained in this way is not exactly solvable, we make an expansion in the parameters $\a$ and $\b$.
In one dimension, eqs.~(3) become at leading order
$$\hx_\m=x_\m-{\b\over 4}(x_\m p^2+p^2x_\m),\qquad \hp_\m=p_\m-{\a\over 4}(p_\m x^2+x^2p_\m),\eqno(13)$$
where we have chosen a Hermitian realization.

One can now work either in position or momentum representation. In the first case,
$$x_\m\to x_\m,\quad p_\m\to-i{\de\over\de x_\m}.\eqno(14)$$
Although the choice of a Hilbert space for the exact theory is difficult, in the context of our approximation we may take it as the space of square-integrable
functions of $x$ in the interval $|x|\le1/\sqrt\a$, with standard scalar product. At leading order this is consistent with the requirement that
$|\hx|\le1/\sqrt\a$, which is natural in our representation.

The one-dimensional \schr equation for a free particle thus becomes
$$(\hp^2-2mE)\,\q=-(1-\a x^2)\,{d^2\q\over dx^2}+2\a x{d\q\over dx}+\({\a\over2}-2mE\)\q=0.\eqno(15)$$
Requiring that $\q$ and its derivative are regular at $x=\pm1/\sqrt\a$, the solution is given by
$$\q\propto P_n(\sqrt\a\,x),\eqno(16)$$
with $P_n$ Legendre polynomials with integer index $n$ such that $n(n+1)={2mE\over\a}-\ha$.
Hence, the energy has a discrete spectrum
$$E_n={\a\over2m}\(n^2+n+\ha\).\eqno(17)$$
In this approximation, the energy does not depend on $\b$.
Notice that $E={p^2\over2m}$ must be less than ${1\over2m\b}$, and hence the quantum number $n$ has an upper bound $n\le\ha\(\sqrt{{2\over\a\b}-1}-1\)$,
which is however extremely large.
\medskip

In the momentum \rep we have instead,
$$p_\m\to p_\m,\quad x_\m\to i{\de\over\de p_\m},\eqno(18)$$
and the \schr equation becomes
$$(\hp^2-2mE)\,\q=\a p^2{d^2\q\over d p^2}+2\a p\,{d\q\over dp}+\(p^2-2mE+{\a\over2}\)\q=0.\eqno(19)$$
The regular solution is given by
$$\q\propto j_n\({p\over\sqrt\a}\),\eqno(20)$$
with $j_n$ a spherical Bessel function with index $n$ such that $n(n+1)={2mE\over\a}-\ha$,
recovering the previous result.

\bigskip
A less trivial example is given by the harmonic oscillator, with Hamiltonian
$$H={\hp^2\over2m}+{m\o^2\hx^2\over2}.\eqno(21)$$
In position representation, the \schr equation can be written as
$$\Big(1-A^2x^2\Big)\,{d^2\q\over dx^2}-2A^2x{d\q\over dx}+\(2mE-{A^2\over2}-m^2\o^2 x^2\)\q=0,\eqno(22)$$
with
$$A^2=\a+\b m^2\o^2.\eqno(23)$$
Changing variable to $z=Ax$, one obtains a special case of spheroidal wave equation [22],
$$(1-z^2)\,{d^2\q\over dz^2}-2z{d\q\over dz}+\({2mE\over A^2}-\ha-{m^2\o^2\over A^4}z^2\)\q=0,\eqno(24)$$
whose solutions are given in terms of spheroidal wave functions $S_{0n}$, defined by means of a three-term recursion relation [22],
$$\q_n\propto S_{0n}(\g,Ax),\eqno(25)$$
with $\g={m\o\over A^2}$.
Notice that eq.~(24) has not a smooth limit for $A\to0$.
From (24) it follows that
$$E_n={A^2\over2m}\(\l_n+\ha\).\eqno(26)$$

The eigenvalues $\l_n$ are discrete, but do not have a simple expression. They can be obtained from a continued fraction
and can be written as a power series in $\o^2$. They are discussed in several texts, see for example [23].
One has
$$\l_n=\sum_{k=0}^\inf l_{2k}\g^{2k},\eqno(27)$$
and
$$l_0=n(n+1),\qquad l_2=\ha\(1-{1\over(2n-1)(2n+3)}\),\qquad{\rm etc.} \eqno(28)$$

However, since $\g$ is assumed to be very large, we are more interested in their asymptotic expansion, which is given by
$$\l_n=2\(n+\ha\)\g-{1\over2}\(n^2+n+{3\over2}\)+o\({1\over\g}\),\eqno(29)$$
so that
$$E_n=\(n+\ha\)\o-{A^2\over4m}\(n^2+n+{1\over2}\)+o\({A^4\over m^2\o}\).\eqno(30)$$
Besides the standard term, one has corrections proportional to powers of $A^2$. Also this result is analogous to TSR [18],
where $E_n=\(n+\ha\)\o+{A^2\over2m}\(n^2+n+\ha\)+o\({A^4\over m^2\o}\)$, but the coefficients of the correction terms are different,
and in particular have opposite sign.

The results of this section can be analytically continued to negative values of $\a$ and $\b$, taking into account that the \bc must
be chosen differently in these cases.

\bigbreak
\section{5. Final remarks}
We have considered the quantum mechanics of the one-dimensional Yang model in its simplest realization on phase space.
The investigation of this simple case is relevant because, as shown in similar situations [15,18], the results are qualitatively similar
to those valid in three dimensions.
The results can be compared with those obtained for analogous models, like TSR or models derived from the generalized uncertainty principle.
A greater variety of possibilities arises in the present case for the modifications of the \ur and for the spectrum of the harmonic oscillator,
because of the freedom to choose independently the signs of the coupling constants. As already shown in [10] and [5,18], in fact, changing the
signs of the couplings in the deformed \ur greatly modifies the physics, leading for example to the disappearance of the minimal length.

From the results obtained, the most interesting case in our model seems to be that of negative deformation constants, since both minimal length
and momentum arise.
In that case, the energy of the free particle is not quantized, and the correction to the harmonic oscillator spectrum have positive sign.

In this paper we have not been mathematically rigorous. A more thorough investigation would require the definition of a suitable Hilbert
space and the study of the functional analysis of the position and momentum operators, in particular when minimal uncertainties are present,
like in the similar case of two-parameter deformation of the uncertainty principle studied in [12].
This analysis is more involved and we plan to afford it in a forthcoming paper.

Further investigations are also required to understand possible phenomenological implications of our results. For example, their relevance
for statistical mechanics has been studied in [7]. Also applications to molecular dynamics in analogy with ref.~[16] can be investigated.

\bigbreak
\beginref

\ref [1] C.N. Yang, \PR{72}, 874 (1947).

\ref [2] H.S. Snyder, \PR{71}, 38 (1947).

\ref [3] M. Born, \RMP{21} 463 (1949).

\ref [4] C. Chryssomakolos and E. Okon, \IJMP{D13}, 1817 (2004).

\ref [5] S. Mignemi, \PR{D84}, 025021 (2011).

\ref [6] V.V. Khruschev, A.N. Leznov, Grav. Cosmol. {\bf 9}, 159 (2003).

H.G. Guo, C.G. Huang and H.T. Wu, \PL{B663}, 270 (2008).

S. Meljanac and S. Mignemi, \PL{B833}, 137289 (2022).

S. Meljanac and S. Mignemi, \JMP{64}, 023505 (2023).

T. Martini\'c-Bila\'c,  S. Meljanac and S. Mignemi, \JMP{64}, 122302 (2023).

J. Lukierski, S. Meljanac, S. Mignemi, A. Pachol and M. Woronowicz, \PL{B854}, 138729 (2024).

\ref [7] A. Pachol, \NP{B1010}, 116771 (2025).

\ref [8] J.J. Heckman and H. Verlinde, \NP{B894}, 58 (2015).

D. Roumelioti, S. Stefas, G. Zoupanos, \arx{2407.07044}.

\ref [9] G. Veneziano, Europhys. Lett. {\bf2}, 199 (1986).

M. Maggiore, \PL{B304}, 63 (1993).

M. Maggiore, \PL{B319}, 83 (1993).

R.J. Adler and D.I. Santiago, \MPL{A14}, 1371 (1999).

F. Scardigli, \PL{B452}, 39 (1999).

\ref [10] P. Jizba, H. Kleinert and F. Scardigli, \PR{D81}, 084030  (2010).

\ref [11] B. Bolen and M. Cavagli\`a, \GRG{37}, 1255 (2005).

C. Bambi and F.R. Urban, \CQG{25}, 095006 (2008).

S. Mignemi, \MPL{A25}, 1697 (2010).

\ref [12] A. Kempf, \JMP{35}, 4483 (1995).

H. Hinrichsen, A. Kempf, \JMP{37}, 2121 (1996).

\ref [13] M.I. Park, \PL{B659}, 698 (2008).

\ref [14] A. Kempf, G. Mangano and R.B. Mann, \PR{D52}, 1108 (1995).

\ref [15] F. Brau, \JoP{A32}, 7691 (1999).

L.N. Chang, D. Minic, N. Okamura and T. Takeuchi, \PR{D65}, 125027  (2002).

\ref [16] A. Maireche, Rev.\ Mex.\ Fis. {\bf 71}, 020401 (2025).

A. Maireche, Jordan Journal of Physics {\bf 17}, 587 (2024).

A. Maireche, Int.\ J.\ Geom.\ Meth.\ Mod.\ Phys.\ {\bf 21}, 2450129 (2024).

\ref [17] J. Kowalski-Glikman and L. Smolin, \PR{D70}, 065020 (2004).

\ref [18] S. Mignemi, \CQG{29}, 215019 (2012).

\ref [19] S. Meljanac and S. Mignemi, \IJMP{A38}, 2350182 (2023).

T. Martini\'c-Bila\'c,  S. Meljanac and S. Mignemi, SIGMA {\bf 20}, 049 (2024).

\ref [20] A. Ballesteros, G. Gubitosi and F.J. Herranz, \CQG{37}, 195021 (2020).

G. Gubitosi and S. Mignemi, Universe {\bf 8}, 108 (2022).

\ref [21] S. Detournay, C. Gabriel and P. Spindel, \PR{D66}, 125004 (2002).

\ref [22] A.H. Wilson, \PRS{A118}, 617 (1928).

\ref [23] M. Abramowitz and I.A. Stegun, {\sl Handbook of mathematical fuctions}, Dover 1965.
\endref
\end